\documentclass[12pt]{article}

\usepackage{graphicx}

\title{\bf A study of the pion effective mass at finite temperature using the linear sigma model
\footnote{Contribution to the proceedings of the conference: X Hadron Physics and VII Relativistic Aspects of Nuclear Physics: A joint
meeting on QCD and QGP, (HADRON-RANP 2004), 28 March--3 April 2004, Angra dos Reis, Rio de Janeiro, Brazil.}}

\author{\bf Nicholas Petropoulos{\footnote{email:nicholas@teor.fis.uc.pt}}\\
\\
\it Centro de F\'\i sica Te\'orica, Departamento de F\'\i sica\\
\it Universidade de Coimbra, P-3004-516 Coimbra, Portugal}
\date{16 June 2004}

\begin{document}
\maketitle

\begin{abstract}
We have used the Cornwall--Jackiw--Tomboulis method of composite operators 
and its formulation at finite temperature by Amelino-Camelia and Pi, in order to calculate the effective potential
of the $O(4)$ linear sigma model beyond the Hartree approximation. We have obtained a system of gap equations for the effective 
mass of sigma and the pions as well as for the condensate, the order parameter of the chiral phase transition. We find that
the thermal effective mass of the pions at low temperatures 
remains lower than in the Hartree case, nevertheless deviates
from the chiral limit. Our observation is consistent with other results which 
have been published previously. 
\end{abstract}

\newpage


The linear sigma model is very often used as an effective theory to QCD in order to obtain insight into  the nature
of the chiral phase transition. Among the quantities of interest is the velocity of pions in medium as well as the 
pion dispersion relation and pion dissipative properties. This is a subject which has attracted some attention 
and has been studied in a variety of models, including the Skyrme model \cite{Lee:2003rj}, the 
linear sigma model \cite{Rischke:1998qy,Son:2002ci,Pisarski:1996mt,Ayala:2002qy} as well as 
chiral perturbation theory \cite{Song:1994ip,Shuryak:1990ie,Schenk:1993ru,Goity:1989gs}.

The essential physics is included in the energy expression  of the pions  propagating in a thermal bath
\begin{equation}
p_0^2={\rm \bf p^2}+ m_\pi^2 + \Sigma_\pi(p_0,{\rm\bf p})~,
\label{Eq:Selfenergy}
\end{equation} 
where $\Sigma_\pi(p_0,{\rm\bf p})$ is the pion self--energy
and depends strongly on the physical conditions of the medium
in which the pion propagates. When the medium is in thermal equilibrium, the self--energy is determined 
by the temperature. The self--energy contains a real part which is related to 
pion velocity and dispersion, while the imaginary part encodes the information about the pion absorption in the medium where
it propagates.

\begin{figure}
  \includegraphics[scale=0.65]{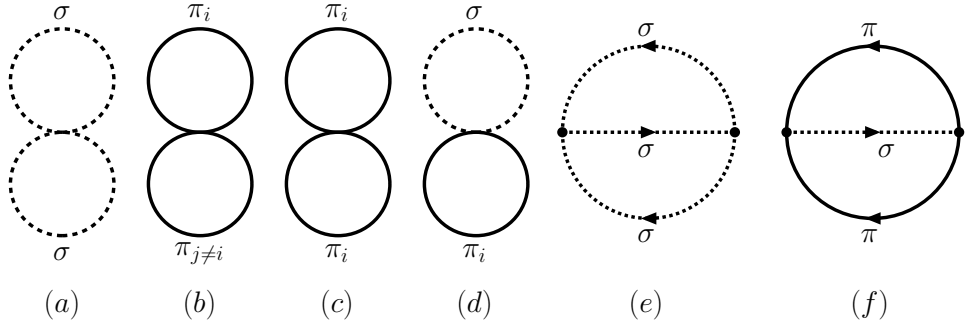}
  \caption{The set of 2PI graphs which contribute to the thermal effective potential of the O(4) linear sigma 
model up to two loops.  The graphs a-d are usually called double--bubble while e,f sunset diagrams.}
  \label{Fig:Fig1}
\end{figure}

We use the $O(4)$ version  of the linear sigma model with the Lagrangian given by
\begin{equation} 
\mathcal{L}=\frac{1}{2}(\partial_{\mu}\Phi)^2- 
\frac{1}{2}m^2 \Phi^2 
-\frac{1}{4!}\lambda (\Phi^2)^2-\varepsilon\sigma~,
\label{Eq:Lagrangian-N}  
\end{equation}
where $\Phi=(\sigma,\pi_1,\pi_2,\pi_3)$, while for $\varepsilon \neq 0$ we deviate from the chiral limit
and the pions are massive. In order to study the propagation properties of the pions at finite 
temperature, we have obtained 
an expression for the thermal effective potential of the linear sigma model up to two loops \cite{Petropoulos:2004bt}. 
As it is shown 
in Fig.~\ref{Fig:Fig1}, there are two types 
of graphs which contribute to the effective potential up to this level.


We have used the  method of composite operators which is  a nice way to  perform systematic selective summations in the loop expansion
of the effective action and the effective potential. In this case, the effective action $\Gamma(\phi,G)$ is the generating  
functional of the two--particle irreducible (2PI) vacuum graphs and depends 
on the constant background field $\phi(x)$,  as well as on the dressed propagators  $G(x,y)$. This formalism was 
initially used by Cornwall--Jackiw--Tomboulis (CJT) \cite{Cornwall:1974vz} for a study of the $O(N)$ model at zero 
temperature, but it has been generalized to finite temperature  
by Amelino-Camelia and Pi \cite{Amelino-Camelia:1993nc}, and was used  
for investigations of the effective potential of the  
$\lambda\phi^4$ theory, the linear sigma model and gauge theories. The method of composite operators
was used long ago in condensed matter context by Luttinger and Ward \cite{Luttinger:1960ua}, as well 
as by Baym \cite{Baym:1962sx} in self--consistent 
approximations of many body systems. Nowadays variations of the technique are  very popular and it has many applications in systems 
at finite temperature in or out of equilibrium as discussed in detail in the contributions of Berges 
and Blaizot in these proceedings.

According to CJT method, the functional derivatives of the effective potential with respect to dressed propagators and 
the background field result in a system of gap equations 
for the thermal effective masses of pions and sigma as well as for the condensate. We have examined the 
problem in two limiting cases. First in the Hartree approximation where we have taken into account only the graphs 
(a)-(d) in  Fig.~\ref{Fig:Fig1}. In this case the self-energy only contains a real part. Going beyond Hartree and guided by 
the low energy theorem we have performed  a selective
summation of the relevant graphs so we have considered only the graph (f) in  Fig.~\ref{Fig:Fig1}. Eventually we have 
ended  up with a gap equation 
for the thermal effective mass of the pions valid only
in the low temperature region. For this calculation  we have made the assumption that the sigma mass varies little
with temperature. We were able this way to obtain the contribution to the thermal pion mass due to the real part
of the pion self energy. Details of this study have been 
presented in \cite{Petropoulos:2004bt}, while the Hartree case is discussed in \cite{Petropoulos:1998gt}
as well as by Rischke and Lenaghan \cite{Lenaghan:1999si}.

\begin{figure} 
\includegraphics[scale=0.79]{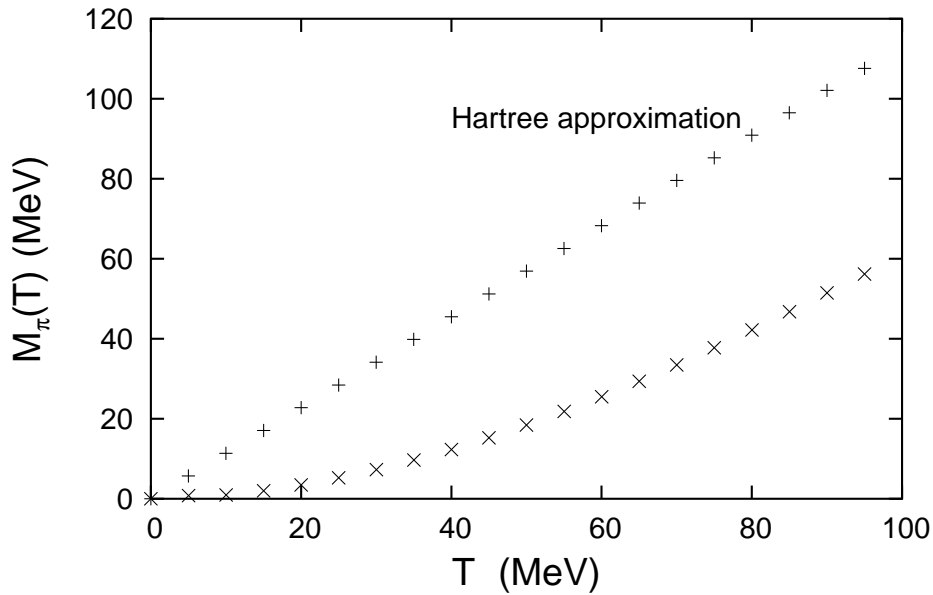} 
\caption{The thermal effective pion mass as a function of temperature.}
\label{Fig:Fig2}
\end{figure}

Our conclusion is reflected in Fig.~\ref{Fig:Fig2}. There we can see that even if we 
have started with massless pions ($\varepsilon=0$), we find that in the 
Hartree approximation, the 
thermal pion mass grows linearly
with temperature \cite{Petropoulos:1998gt,Lenaghan:1999si}. On the contrary, taking into 
account the sunset diagram (Fig.~\ref{Fig:Fig1}f), we find that the thermal mass 
depends on temperature quadratically \cite{Petropoulos:2004bt}. This of course, at low temperature, 
keeps the pions closer to their Goldstone nature. Our observation is consistent with earlier results 
by Itoyama--Mueller \cite{Itoyama:1983up}
and Pisarski--Tygtat \cite{Pisarski:1996mt} using the linear sigma model, as well as with other investigations  using 
the techniques of chiral perturbation theory \cite{Toublan:1997rr}.

The author would like to thank and congratulate the organisers for the excellent  meeting. 
It has been a pleasure
to exchange ideas and receive advise from Heron Caldas, Marcelo Hott, \'Agnes M\'ocsy, Marina Nielsen, D\'ebora Menezes,
Cristian Villavicencio, Marcelo Loewe, Thomas Ullrich,
Cristof Wetterich, Jurgen Berges, L\'aszl\'o Csernai, Ulrich Heinz and Larry McLerran. A critical reading of the manuscript by
Constan\c{c}a  Provid\^encia, and lengthy discussions with Eef van Beveren on the 
nature of scalar mesons and especially the sigma meson are  really appreciated. This work is supported by
the {\it Funda\c{c}\~ao para Ci\^encia e a Tecnologia} of the {\it Minist\'erio da Ci\^encia e da Tecnologia} of Portugal under contract
POCTI/FNU/49555/2002.

\end{document}